\documentclass[epj]{svjour}
\usepackage{latexsym}
\usepackage{graphicx}
\usepackage{multirow}
\usepackage{color}
\usepackage{amssymb}
\begin{document}
\title{Do hyperons exist in the interior of neutron stars ?}
\author{Debarati Chatterjee\inst{1,2} \and Isaac Vida\~na\inst{3}
}                     
\offprints{dchatterjee@lpccaen.in2p3.fr}  
\offprints{ividana@fis.uc.pt}         

\institute{LUTH, Observatoire de Paris, CNRS, Universit\'e Paris Diderot, 5 place Jules Janssen, F-92190 Meudon, France 
\and Laboratoire de Physique Corpusculaire, ENSICAEN, 6 Boulevard Mar\'echal Juin, F-14050 Caen C\'edex, France
\and CFisUC, Department of Physics, University of Coimbra, rua Larga s/n, PT-3004-516 Coimbra, Portugal}
\date{Received: date / Revised version: date}
%
\abstract{In this work we review the role of hyperons on the properties of neutron and proto-neutron stars.  In particular, we revise the so-called ``hyperon puzzle", go over some of the solutions proposed to tackle it, and discuss the implications that the recent measurements of unusually high neutron star masses have on our present knowledge of hypernuclear physics. We reexamine also the role of hyperons on the cooling properties of newly born neutron stars and on the so-called r-mode instability.
} 
\maketitle
%
\section{Introduction} 
\label{intro}

Strangeness adds a new dimension to the evolving picture of nuclear physics giving us an opportunity to study fundamental interactions from an enlarged perspective. The presence of strangeness in finite nuclear systems ({\it i.e.,} hypernuclei) is well established experimentally, and it constitutes a unique probe of the deep nuclear interior which makes it possible to study a variety of otherwise inaccessible nuclear phenomena and, thereby, to test nuclear models. 

Furthermore, there is a growing evidence that strange particles can have significant implications for astrophysics. In particular, the presence of strangeness, both in a confined (hyperons, kaons) or deconfined (strange quark matter) form, in the dense inner core of neutron stars is expected to have important consequences for the equation of state (EoS), composition, structure and evolution of these objects. However, 
other than the proposed detection of gravitational waves (GW), there is no direct method yet to probe the internal composition of neutron stars, and one must rely on theoretical models of the dense matter EoS to relate neutron star observations with possible signatures of strangeness in their core. These models are usually constrained by nuclear 
and hypernuclear data at densities close to the normal nuclear saturation density ($\rho_0\sim 0.16$ fm$^{-3}$) and
low values of the isospin asymmetry. However, one of the main drawbacks of these models is the lack of experimental data to constrain their parameters at high densities and large isospin asymmetries. Neutron star observables can provide these additional constraints and be used to check the validity of many of the large number of exiting models ruling out all those which are incompatible with the observational data. Unfortunately, the interpretation of such data is highly model dependent and presents large uncertainties. Nevertheless, the masses of neutron star binaries, determined using post-Keplerian parameters, are among the most accurate measurements. Until very recently, 
the most precise neutron star mass observations clustered around a ``canonical" value of $1.4-1.5M_\odot$  \cite{hulsetaylor,hulsetaylor2,thorset99}, and the only requirement to be fulfilled by any reliable EoS was to predict a maximum neutron star mass, $M_{max}$, compatible with this value. Most of the existing models, including those with exotic components, such as hyperons, meson condensates or deconfined quark matter, predict $M_{max} \geq 1.4-1.5M_\odot$. Therefore, this constraint is not stringent enough to discriminate between the different models. Fortunately, a breakthrough came recently with the precise measurement of the unusually high masses of the millisecond pulsars 
PSR J1903+0327 ($1.667 \pm 0.021M_\odot$) \cite{Freire,Freire2,Freire3}, PSR J1614-2230 ($1.97 \pm 0.04M_\odot$) \cite{Demorest} and PSR J0348+0432 ($2.01 \pm 0.04M_\odot$) \cite{Antoniadis}. These new measurements, particularly the
last two, impose a stronger constraint on the models of the dense matter EoS implying, now, that any reliable one should predict $M_{max} \geq 2M_\odot$. 

It is known that any additional degree of freedom ({\it e.g.,} hyperons) appearing in the neutron star interior softens its EoS and reduces its mass. Therefore, in view of this new and severe observational constraint a natural question arises: can hyperons, or strangeness in general, still be present in the neutron star interior if $M_{max}$ is reduced to values smaller than $2M_\odot$, although their presence is energetically favorable? This question is at the origin of the so-called ``hyperon puzzle", whose solution is not easy and it is presently a subject of very active research. In this work we go over some of the solutions that have been proposed to tackle this problem, and discuss the implications that the discovery of massive neutron star have on our present knowledge of hypernuclear physics.

In addition to the maximum mass, the measurement of radii can also serve to constrain the EoS and estimate the strangeness content in neutron stars. In a recent work, for instance, Provid\^{e}ncia and Rabhi \cite{Providencia2013} have shown that, for a given mass, the radius of the star decreases linearly with the increase of the total hyperon content, and have estimated an upper limit of the expected hyperon fraction in neutron stars from radius determinations. Fortin {\it et al.} \cite{Fortin} have also shown that the observational constraint on the maximum mass implies that the radii of hyperonic stars with masses in the range $1-1.6M_\odot$ must be larger than $13$ km due to a pre-hyperon stiffening required for the EoS. However, despite the theoretical effort, the analysis of present observations from quiescent low-mass X-ray binaries is still controversial. Whereas the one of Steiner {\it et al.} \cite{Bayesian,Bayesian2} indicates neutron star radii in the range of $10.4-12.9$ km, that of Guillot {\it et al.} \cite{Guillot,Guillot2} points towards smaller radii of $\sim 10$ km or less, which in combination with the heavy mass measurements implies a stiff EoS on the very edge of causality \cite{Dexheimer1411}. If the result of Guillot {\it et al.,} is confirmed by further analysis then the simultaneous existence of massive neutron stars and objects with small radii would be a very complicated problem to solve  
for any of the existing models of the pure nucleonic EoS. A solution to this problem that has been proposed
is the possible existence of the so-called ``twin stars", stars with similar masses but smaller radii than those 
made only of nucleons. Recently, it has been conjectured that these twin stars could in fact be composed of strange hadronic or quark matter \cite{Dexheimer1411,Drago1309}. The interested reader is referred to these works for
a detailed discussion of this interesting astrophysical scenario.

Other neutron star properties, such as their thermal and structural evolution, can also be very sensitive to the composition and, therefore, to the strangeness content of their interior. In particular, the cooling of neutron stars may be affected by the presence of strangeness, which can modify neutrino emissivities allowing for fast and more efficient cooling mechanisms. Measurements of the surface temperatures of neutron stars, with satellite-base X-ray observatories, can help us to determine and quantify the role of the strange components of dense matter in the cooling properties of these objects. This is the case, for instance, of the compact star in Cassiopeia A (Cas A) \cite{casA,casA2} whose thermal evolution, studied over the last 10 years, seems to indicate the necessity of fast cooling mechanisms to explain the data and, therefore, it can be used to constrain different EoSs with strangeness content. In Ref.\ \cite{Sedrakian2013}, for instance, it has been suggested that such fast cooling in Cas A could proceed via phase transitions among different superconducting phases of quark matter instead of less exotic Cooper pair-breaking processes. Cooling studies of other neutron stars can serve in general to rule out slow cooling by invoking  Urca 
(direct and modified) processes involving exotic matter, and help up to constrain the maximum strangeness fraction in the neutron star core.

Finally, the emission of GW in hot and rapidly rotating newly born neutron stars due to the so called r-mode instability \cite{rmode,rmode2} can also be affected by the presence of strangeness, because bulk viscosity of neutron star matter is dominated by the contribution of the strange components as soon as they appear in the neutron star interior.  

As said in the abstract, in this work we review the role of hyperons on the properties of neutron stars. Before, however, in the next section we briefly revise some of the laboratory constraints on the dense matter EoS with strangeness derived from hypernuclear research and heavy-ion collisions. Then, in section
\ref{sec:hyp_ns}, we present some of the ideas proposed to solve the hyperon puzzle. In section \ref{sec:qm_ns} we analyse the role of quark matter on massive neutron stars, whereas in sections \ref{sec:cool} and \ref{sec:rmode} we revise, respectively, the effect of hyperons on the cooling and r-mode instability. Finally, in section \ref{sec:summary}, we
present a summary of the present work and future perspectives on this area of research.

\section{Laboratory constraints to the nuclear EoS with strangeness degrees of freedom}
\label{sec:hyppys}

One of the goals of hypernuclear physics is to relate hypernuclear observables with the bare hyperon-nucleon
(YN) and hyperon-hyperon (YY) interactions. Contrary to the nucleon-nucleon (NN) interaction, which is fairly well known due to the larger number of existing scattering data and measured properties of nuclei, YN and YY interactions are still poorly constrained. The experimental difficulties associated with the short lifetime of hyperons together with the low intensity beam fluxes have limited the number of $\Lambda$N and $\Sigma$N scattering events to approximately 600 \cite{engelmann66,alexander68,sechi68,kadyk71,eisele71}, and that of $\Xi$N events to very few. In the case of the YY interaction the situation is even worse because no scattering data exists at all. This limited amount of data is not enough to fully constrain these interactions. A commonly followed approach to construct a general baryon-baryon interaction is to start from a given NN one and to extend it to the strange sector by imposing the SU(3)-flavor symmetry. This has been mainly done within the framework of a meson-exchange theory by the Nijmegen \cite{nijmegen,nijmegen2,nijmegen3,nijmegen4,nijmegen5} and J\"{u}lich \cite{juelich,juelich2} groups although, recently, a new approach based on chiral perturbation theory has emerged as a powerful tool \cite{chiral,chiral2}.

In the absence of scattering data, alternative information on the YN and YY interactions can be obtained from the study of hypernuclei, bound systems composed of nucleons and one or more hyperons. They were first observed in 1952 with the discovery of a hyperfragment by Danysz and Pniewski in a ballon-flown emulsion stack \cite{danpni}. Since then the use of high-energy accelerators as well as modern electronic counters have led to the identification of more than 40 single $\Lambda$-hypernuclei and few double-$\Lambda$ ones. The existence of single 
$\Sigma$-hypernuclei has not been experimentally confirmed yet without any ambiguity, suggesting this that the $\Sigma$N interaction is probably repulsive.

Single $\Lambda$-hypernuclei can be produced by several mechanisms such as:

\vspace{0.25cm}
{\it Strangeness exchange reactions $(K^-,\pi^-)$}: 

\begin{equation}
K^-\,\, + \,\, ^AZ \,\, \rightarrow\,\,  ^A_{\Lambda}Z\,\,+\,\,\pi^- \ ,
\end{equation}
where a neutron hit by a $K^-$ is changed into a $\Lambda$ hyperon emitting a $\pi^-$. These experiments measure mainly the hyperon binding energies and allow the identification of excited hypernuclear levels. The hypernuclear mass, from which the hyperon binding energy can be deduced as
\begin{equation}
B_{^A_{\Lambda}Z}=B_{^AZ}+M_{^AZ}+M_{\Lambda}-M_{N}-M_{^A_{\Lambda}Z} \ ,
\end{equation}
is obtained as follows
 \begin{equation}
M_{^A_{\Lambda}Z}=\sqrt{\left(E_{\pi^-}-E_{K^-}-M_{^AZ}\right)^2+\left(\vec p_{\pi^-}-\vec p_{K^-}\right)^2} \ .
\end{equation}
Two magnetic spectrometers, to measure the incident $K^-$ momentum and the outgoing $\pi^-$ one,
both with good energy resolution are required in order to achieve a good hypernuclear mass resolution.

In some cases a $K^-$ beam, with rather low-momentum, is injected on thick nuclear targets that stop the $K^-$ before it decays. The $K^-$ losses its energy in the target, and is eventually trapped in atomic orbits of a kaonic atom through various atomic processes. The $K^-$ is absorbed in the final state by the atomic nucleus. The kaon capture proceeds mainly with the emission of a pion and the formation of a hypernucleus
\begin{equation}
K^-_{stopped}\,\, + \,\, ^AZ \,\, \rightarrow\,\,  ^A_{\Lambda}Z\,\,+\,\,\pi^- \ .
\end{equation}
Since this reaction occurs essentially at rest, in this case it is necessary to measure only the momentum of the emitted pion in order to determine the hypernuclear mass
\begin{equation}
M_{^A_{\Lambda}Z}=\sqrt{\left(E_{\pi^-}-E_{K^-}-M_{^AZ}\right)^2+\vec p_{\pi^-}^2} \ .
\end{equation}
Therefore, one magnetic spectrometer is enough. These reactions, initially carried out at CERN, have
been studied mainly at BNL in the USA, and at KEK and J-PARC in Japan.

\vspace{0.25cm}
{\it Associate production reactions $(\pi^+,K^+)$}: 

\begin{equation}
\pi^+\,\, + \,\, ^AZ \,\, \rightarrow\,\,  ^A_{\Lambda}Z\,\,+\,\,K^+ \ .
\end{equation}

Here, an $s\bar{}s$ pair is created from the vacuum and a $K^+$ and a $\Lambda$ are produced in the final state (the so-called associate production). The production cross section is reduced, compared to the one of the strangeness exchange reaction. However, this drawback is compensated by the greater intensities of the $\pi^+$ beams. The hypernuclear mass is obtained by measuring the $\pi^+$ and the $K^+$ momenta with two spectrometers as in the case of the $(K^-,\pi^-)$ reaction. These experiments have been also performed at BNL and KEK, and latter at GSI (Germany).

\vspace{0.25cm}
{\it Electro-production reactions $(e,e'K^+)$}: 

\begin{equation}
e^-\,\, + \,\, ^AZ \,\, \rightarrow\,\, e^-\,\,+\,\, K^+\,\,+\,\,  ^A_{\Lambda}(Z-1) \ .
\end{equation}

This process provides a high-precision tool to study $\Lambda$-hypernuclear spectroscopy, with energy resolutions of several hundred keV \cite{hugenford94}. At present only two laboratories in the world, the JLAB (USA) and MAMI-C (Germany), have the instrumental capabilities to perform experiments on hypernuclear spectroscopy by using electron beams. The electron beams have excellent spatial and energy resolutions, so this reaction is used for studies of hypernuclear structure. The electro-production of hypernuclei can be well described by a first order perturbation calculation as the exchange of  a virtual photon between the electron and a proton of the nucleus which is changed into a $\Lambda$ hyperon. Although the cross section for this reaction is about 2 orders of magnitude smaller than that of the $(\pi^+,K^+)$ one, this can be compensated by larger electron beam intensities. Since the cross section falls rapidly with increasing transfer momentum, and the virtual photon flux is maximized for an electron scattering angle near zero degrees, experiments must be done within a small angle around the direction of the virtual photon. The experimental geometry requires two spectrometers, one to detect the scattered electrons which defines the virtual photons, and one to detect the kaons. Both of these spectrometers must be placed at extremely forward angles. Because of this, a magnet is necessary to deflect the electrons away from zero degrees into their respective spectrometers. In addition, since many pions, positrons and protons are transmitted through the kaon spectrometer, it is required an excellent particle identification, not only in the hardware trigger, but also in the data analysis. By measuring the type of outgoing particles and their energies $(E_{e'}, E_{K^+})$, and knowing the energy of the in-coming electron $(E_e)$, it is possible to calculate the energy which is left inside the nucleus in each event:  
\begin{equation}
E_x=E_e-E_{e'}-E_{k^+} \ ,
\end{equation}
from which it can be deduced the binding energy of the produced hypernuclei.

\begin{figure}[b]
\begin{center}
\resizebox{0.30\textwidth}{!}
{%
\includegraphics[clip=true]{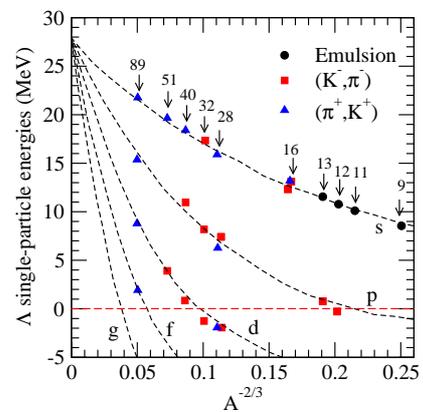}
}
\caption{(Color online) Energy of a $\Lambda$ hyperon in the single-particle orbits s, p, d, f and g of several hypernuclei as a function of $A^{-2/3}$ deduced from emulsion, $(K^-,\pi^-)$ and $(\pi^+,K^+)$ reactions. The dashed lines are drawn just to help the reader.} 
\label{f:fig0}       
\end{center}
\end{figure}

Hypernuclei can be produced in excited states if a nucleon in a p or higher shell is replaced by a hyperon. The energy of this excited states can be released either by emitting nucleons, or, sometimes, when the hyperon moves to lower energy states, by the emission of $\gamma$-rays. Measurements of $\gamma$-ray transitions in $\Lambda$-hypernuclei has allowed to analyze excited levels with an excellent energy resolution. The energy of a $\Lambda$ hyperon in the single-particle orbits s, p, d, f and g of several hypernuclei, deduced from emulsion, $(K^-,\pi^-)$ and $(\pi^+,K^+)$ reactions, is shown as a function of the mass number to the power $-2/3$ in Fig.\ \ref{f:fig0}. The value of $\sim 28$ MeV extrapolated at $A^{-2/3}=0$ is usually interpreted as the binding energy of a single $\Lambda$ hyperon in infinite nuclear matter at $\rho_0$, and it is used to fix the parameters of the majority of the models of the hyperonic matter EoS. Systematic studies of $\Lambda$-hypernuclei indicate the $\Lambda$N interaction is clearly attractive \cite{hashimoto06}.

$\Sigma$-hypernuclei can also be produced by the mechanisms just described. However, as mentioned before, there is not yet an unambiguous experimental confirmation of their existence. 

Double-$\Lambda$ hypernuclei cannot be produced in a single step. To produce them, first it is necessary to create a $\Xi^-$ through the reaction
\begin{equation}
K^-\,\,+\,\,p\,\,\rightarrow\,\,\Xi^-\,\,+\,\,K^+ \ ,
\label{eq:xi1}
\end{equation}
or
\begin{equation}
p\,\,+\,\,\bar p\,\,\rightarrow\,\,\Xi^-\,\,+\,\,\bar \Xi^+ \ .
\label{eq:xi2}
\end{equation}
The $\Xi^-$ should be then captured in an atomic orbit and interact with the nuclear core producing two $\Lambda$ hyperons via a process like {\it e.g.,}
\begin{equation}
\Xi^-\,\,+\,\,p\,\,\rightarrow\,\,\Lambda\,\,+\,\,\Lambda\,\,+\,\,28.5\,\,\mbox{MeV}  \ .
\end{equation}
This reaction provides about $30$ MeV of energy that is equally shared between the two $\Lambda$'s in most cases, leading to the scape of one or both hyperons from the nucleus. We note that $\Xi$-hypernuclei can be produced by means of the reactions (\ref{eq:xi1}) and (\ref{eq:xi2}).  There are very few $\Xi$-hypernuclei that seems to indicate an attractive $\Xi$-nucleus interaction of the order of about $\sim -14$ MeV.
This result is based on the analysis of the experimental data from production reactions such as $^{12}$C$(K^-,K^+)^{12}_{\Xi^-}$Be \cite{kau}. We should mention here also the very recent observation of
a deeply bound state of the $\Xi^-$-$^{14}$N system with a binding energy of $4.38 \pm 0.25$ MeV by
Nakazawa {\it et al.,} \cite{nakazawa15}. This event provides the first clear evidence of a deeply bound state of this system by an attractive $\Xi$N interaction. Future $\Xi$-hypernuclei experiments are being planned at J-PARC.

Double-strange hypernuclei are nowadays the best systems to investigate the properties of the strangeness $S=-2$ baryon-baryon interaction. Emulsion experiments and subsequent analysis \cite{dl1,dl1b,dl2,dl3,dl4} have reported the formation of a few double-$\Lambda$ hypernuclei: $^6_{\Lambda\Lambda}$He, $^{10}_{\Lambda\Lambda}$Be and $^{13}_{\Lambda\Lambda}$B. The 
$\Lambda\Lambda$ bond energy $\Delta B_{\Lambda\Lambda}$ in double-$\Lambda$ hypernuclei is determined experimentally from the measurement of the binding energies of double- and single-$\Lambda$ hypernuclei as
\begin{equation}
\Delta B_{\Lambda\Lambda}=B_{\Lambda\Lambda}(^A_{\Lambda\Lambda}Z)-2B_{\Lambda}(^{A-1}_{\Lambda}Z) \ .
\end{equation}
From the resulting $\Lambda\Lambda$ binding energies, a reasonably large $\Lambda\Lambda$ bond energy of around 4 to 5 MeV emerged in old analysis, contrary to expectation from SU(3) \cite{nijmegen3}. However, a new $^6_{\Lambda\Lambda}$He candidate having a $\Lambda\Lambda$ bond energy of around 1 MeV  was found in 2001 at KEK \cite{nagara}. 
Unless new experiments for the other double-$\Lambda$ hypernuclei also give lower binding energies in the future, it will remain an open question how to reconcile theoretically the weak $\Lambda\Lambda$ attraction found in $^6_{\Lambda\Lambda}$He with the stronger one suggested by the other double-strange systems. Further experiments are planned in the future at BNL, KEK and J-PARC with $K^-$ beams (Eq.\ (\ref{eq:xi1}) ), and at FAIR (Germany) with protons and antiprotons (Eq.\ (\ref{eq:xi2})).

We should mention here also that the hypothesis that absolute strange quark matter might be stable \cite{sqm,sqm2,sqm3,sqm4,sqm5} has led to an intensive search for the double-strange H-dibaryon, which is equivalent to a bound $\Lambda\Lambda$, $\Sigma\Sigma$ or $\Xi$N system. However, till now there has been no experimental evidence of the existence of the H-dibaryon above or below the $\Lambda\Lambda$ threshold.

Theoretically, hypernuclei can be described in a simple model as an ordinary nuclei with hyperons sitting in the single-particle states of an effective hyperon-nucleus potential derived from the YN and YY interactions. Traditionally, hypernuclei have been reasonably well described by a shell-model picture using $\Lambda$-nucleus potentials of \\Woods-Saxon type that reproduce quite well the measured hypernuclear states of medium to heavy hypernuclei \cite{ws1,ws1b,ws2,ws3}. Non-localities and density dependent effects, included in non-relativistic Hartree--Fock calculations using Skyrme YN interactions \cite{skhf1,skhf2,skhf3,skhf4,skhf4b,skhf5,skhf5b,skhf5c,skhf5d} improved the overall fit of the single-particle binding energies. The properties of hypernuclei have also been studied in a relativistic framework, such as Dirac phenomenology, where the hyperon-nucleus potential has been derived from the nucleon-nucleus one \cite{dirac1,dirac2}, or within the relativistic mean field (RMF) theory \cite{rmf1,rmf2,rmf3,rmf4,rmf5,rmf6,rmf7,rmf8,rmf8b}.

Microscopic hypernuclear structure calculations, which provide the desired link of the hypernuclear observables with the bare YN and YY interactions, are also available. They are based on the construction of effective YN and YY $G$-matrices which are obtained from the bare YN and YY interactions by solving the Bethe--Goldstone equation. In earlier microscopic calculations, Gaussian parametrizations of the $G$-matrices calculated in nuclear matter at an average density were employed \cite{g1,g1b,g1c,g1d}. A $G$-matrix obtained directly in finite nuclei was used to study the single-particle energy levels in various single-$\Lambda$ hypernuclei \cite{g2}. Nuclear matter $G$-matrix were also used as an effective interaction in the calculation of the $^{17}_\Lambda$O
spectrum \cite{g3}. The s- and p-wave $\Lambda$ single-particle properties for a variety of $\Lambda$-hypernuclei, from $^5_\Lambda$He to $^{208}_\Lambda$Pb, were derived in Refs.\ 
\cite{g4,g5} by constructing a finite nucleus YN $G$-matrix form a nuclear matter $G$-matrix. We should mention also the very recently  Quantum Monte Carlo study of single- and double-$\Lambda$ hypernuclei up to a mass number $A=91$ of Ref.\ \cite{qmchyp}. The authors of this work have shown that by accurately refitting the parameters of the $\Lambda$N and $\Lambda$NN  forces their calculation can properly describe the available experimental data over a wide range of hypernuclear masses. 

While nuclear and hypernuclear physics gives us valuable information at low energies, heavy-ion collisions
provide information on compressed baryonic matter at intermediate energies \cite{EMMI}. It is possible to compress nuclear matter to baryon densities up to (2-4)$\rho_0$ at beam energies of 0.2-2.0 AGeV in heavy-ion collisions, which are relevant densities for the neutron star interior where exotic components can appear \cite{Li02}. In addition, investigations using ions with varying Z/N ratios allows the possibility to probe isospin asymmetry of dense matter. Strange hadrons are produced in abundance in heavy-ion collisions at intermediate densities, suggesting that their presence cannot be ignored in dense matter theories. 

The analysis of subthreshold production of kaons in heavy-ion collisions by the KaoS collaboration at GSI using transport simulations indicated that the nuclear EoS at (2-3)$\rho_0$ is soft \cite{Hartnack}. In a recent work by Sagert {\it et al.} \cite{SagertTolos} the implications of this result for the maximum neutron star mass were proved. It was shown that the EoS compatible with the heavy-ion data can be reconciled with the massive neutron star observations if the EoS at higher densities becomes sufficiently stiff. The influence of the critical density for the transition from a soft EoS (compatible with heavy-ion data from KaoS experiment at 2-3 $\rho_0$) to the stiffest possible EoS (limited by causality) on the maximum mass of a neutron star was demonstrated. 

To finish this section we recall that a complementary constraint to the dense matter EoS comes from the analysis of the elliptic flow of isospin asymmetric matter in heavy-ion collisions \cite{Danielewicz}. The flow constraint imposes an upper limit on the pressure as a function of density in symmetric matter. Using these analyses, Danielewicz {\it et al.} eliminated strongly repulsive nuclear
EoSs and weakly repulsive EoSs with phase transitions at densities less than 3$\rho_0$, but not EoSs with a transformation to quark matter at high densities. 

\begin{figure*}[t]
\begin{center}
\resizebox{0.70\textwidth}{!}
{%
\includegraphics[clip=true]{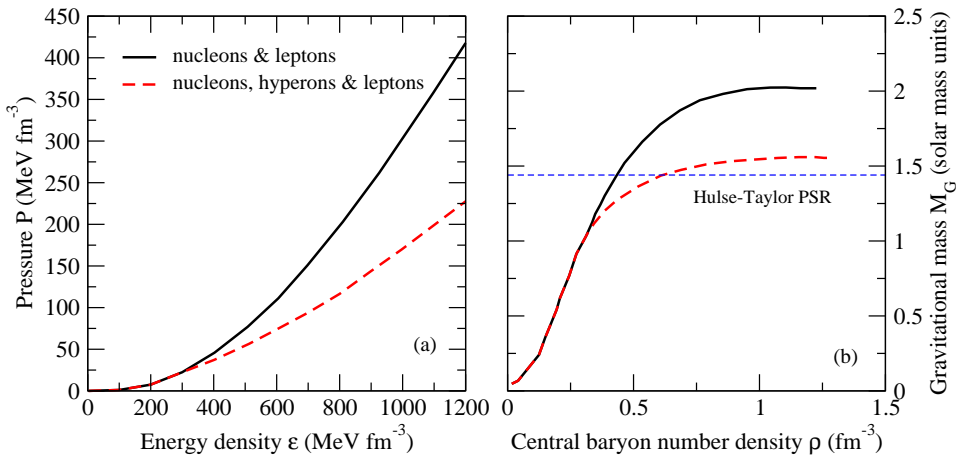}
}
\caption{(Color online) Illustration of the effect of the presence of hyperons on the EoS (panel (a)) and mass of a neutron star (panel (b)). A generic model with (black solid line) and without (red dashed line) hyperons has been considered. The horizontal line shows the observational mass of the Hulse--Taylor \cite{hulsetaylor,hulsetaylor2} pulsar.} 
\label{f:fig1}       
\end{center}
\end{figure*}

\section{Neutron stars and the hyperon puzzle}
\label{sec:hyp_ns}

The composition of the interior of neutron stars is still under debate. Traditionally the core of neutron stars has been modeled as a uniform fluid of neutron-rich nuclear matter in equilibrium with respect to the weak interaction ($\beta$-stable matter). However, with increasing density, new degrees of freedom such as hyperons are expected to appear in addition to nucleons. Contrary to terrestrial conditions, where hyperons are unstable and decay into nucleons through the weak interaction, the equilibrium conditions in neutron stars can make the inverse process happen, so the formation of hyperons becomes energetically favorable. 

The presence of hyperons in neutron stars was considered for the first time in the pioneering work of Ambartsumyan and Saakyan in 1960 \cite{ambsaa60}. Since then, their effects on the properties of these objects have been studied by many authors using either phenomenological 
\cite{skm,skmb,glen,glenb,glenc,glend,rmf,rmfb,rmfc,rmfd,rmfe,rmff,rmfg,rmfh,Hofmann,Huber,Taurines,Gomes2014,Rikovska,Whittenbury,Thomas,Miyatsu,Dhiman07,Dexheimer08,Bednarek2011,Weissenborn1,Weissenborn2,Agrawal,Lopes2014,Oertel2014,Maslov,GuptaArumugam,CharBanik,Typel,Typelb,Colucci,vanDalen,Lim2014} 
or microscopic \cite{micro,micro2,micro3,micro4,micro5,micro6,micro7,micro8,vlowk,dbhf1,dbhf2,dbhf2b,taka,takab,vidanatbf,yamamoto,yamamotob,yamamotoc,lonardoniprl} approaches for the neutron star matter EoS with hyperons.  Phenomenological approaches, relativistic or non-relativistic, are based on effective density-dependent interactions which typically contain a certain number of parameters adjusted to reproduce nuclear and hypernuclear observables, and neutron star properties. Skyrme-type interactions \cite{skm,skmb} and RMF models 
\cite{glen,glenb,glenc,glend,rmf,rmfb,rmfc,rmfd,rmfe,rmff,rmfg,rmfh,Hofmann,Huber,Taurines,Gomes2014,Rikovska,Whittenbury,Thomas,Miyatsu,Dhiman07,Dexheimer08,Bednarek2011,Weissenborn1,Weissenborn2,Agrawal,Lopes2014,Oertel2014,Maslov,GuptaArumugam,CharBanik,Typel,Typelb,Colucci,vanDalen,Lim2014} 
are among the most commonly used ones within this type of approach.
Several authors have derived phenomenological EoSs using density-dependent baryon-baryon interactions based on Skyrme-type forces including hyperonic degrees of freedom \cite{skm,skmb}. The features of most of these EoSs rely on the properties of nuclei for the NN interaction, and on the experimental data from hypernuclei for the YN and YY ones. 
RMF models are based in effective Lagrangian densities where the baryon-baryon interactions are described in terms of meson exchanges. A RMF description of the EoS of dense matter with hyperons, which was done for the first time by Glendenning in the 1980s \cite{glen,glenb,glenc,glend} and later by many others
\cite{glen,glenb,glenc,glend,rmf,rmfb,rmfc,rmfd,rmfe,rmff,rmfg,rmfh,Hofmann,Huber,Taurines,Gomes2014,Rikovska,Whittenbury,Thomas,Miyatsu,Dhiman07,Dexheimer08,Bednarek2011,Weissenborn1,Weissenborn2,Agrawal,Lopes2014,Oertel2014,Maslov,GuptaArumugam,CharBanik,Typel,Typelb,Colucci,vanDalen,Lim2014}, 
is turning out nowadays to be one of the most popular ones. The parameters in this approach are usually fixed by the properties of nuclei and nuclear bulk matter for the nucleonic sector, whereas the coupling constants of the hyperons are fixed by symmetry relations and hypernuclear observables.

Microscopic approaches, on the other hand, are based on realistic two-body baryon-baryon interactions that describe the scattering data in free space. These realistic interactions, as it has been said in the previous section, are based on the meson-exchange \cite{nijmegen,nijmegen2,nijmegen3,nijmegen4,nijmegen5,juelich,juelich2} or chiral perturbation theory \cite{chiral,chiral2}.
In order to obtain the EoS one has to solve then the very complicated many-body problem \cite{mbp}. A great difficulty of this problem lies in the treatment of the repulsive core, which dominates the short-range behavior of the interaction. Although different microscopic many-body methods have been extensively used to the study of nuclear matter, up to our knowledge, only the Brueckner--Hartree--Fock (BHF) approximation \cite{micro,micro2,micro3,micro4,micro5,micro6,micro7,micro8} of the Brueckner--Bethe--Goldstone ((BBG) theory, the $V_{low \> k}$ approach \cite{vlowk}, the Dirac--Brueckner--Hartree--Fock (DBHF) theory \cite{dbhf1,dbhf2,dbhf2b}, and very recently the Auxiliary Field Diffusion Monte Carlo (AFDMC) method \cite{lonardoniprl}, have been extended to the hyperonic sector. 

All these approaches agree that hyperons may appear in the inner core of neutron stars at densities of $\sim 2-3\rho_0$. At such densities, the nucleon chemical potential is large enough to make the conversion of nucleons into hyperons energetically favorable. This conversion relieves the Fermi pressure exerted by the baryons and makes the EoS softer, as it is illustrated in panel (a) of Fig.\ \ref{f:fig1} for a generic model with (black solid line) and without (red dashed line) hyperons. As a consequence (see panel (b)) the mass of the star, and in particular the maximum one, is substantially reduced. In microscopic calculations, the reduction of the maximum mass can be even below the canonical one (see {\it e.g.,} Refs.\ \cite{micro,micro2,micro3,micro4,micro5,micro6,micro7,micro8,vlowk,dbhf1,dbhf2,dbhf2b}). This is not the case, however, of phenomenological calculations for which the maximum mass obtained is still compatible with the canonical value. In fact, most relativistic models including hyperons obtain maximum masses in the range $1.4-1.8M_\odot$ \cite{glen,glenb,glenc,glend,rmf,rmfb,rmfc,rmfd,rmfe,rmff,rmfg,rmfh}. 
However, in some exceptional cases, neutron stars with maximum masses larger than $2M_\odot$ have been obtained. Huber {\it et al.} \cite{Huber}, for instance, constructed neutron star matter EoSs based on the relativistic Hartree- and  Hartree--Fock approximation compatible with hypernuclear data and obtained masses larger than $2M_\odot$ for certain range of the hyperon couplings constrained by the binding energies of hyperons in symmetric nuclear matter. Taurines {\it et al.,} \cite{Taurines} achieved large neutron star masses including hyperons by considering a RMF model with density-dependent couplings. These couplings simulate the effect of many-body forces by incorporating non-linear self-interaction and meson-meson interaction terms for the scalar mesons. Recently, Gomes {\it et al.,} \cite{Gomes2014} has extended this model to include other meson fields, both non-strange and strange, and have succeeded in describing neutron stars compatible with the mass constraint. The authors of Ref.\ \cite{Rikovska} predicted the existence of neutron stars with hyperons and masses in the range $1.9-2.1M_\odot$ using the quark meson coupling (QMC) model. This model is derived at a fundamental level from quarks with adjustable parameters fitted to reproduce nuclear and hypernuclear properties. Recently, Whittenbury {\it et al.} \cite{Whittenbury,Thomas} extended the latest version of this model to include the full tensor treatment of the baryon-vector meson couplings within the Hartree--Fock approximation and showed that the $\rho$N tensor coupling is essential to produce a stiff EoS at high densities while keeping a reasonable value of the incompresibility at saturation. This work complemented that of Miyatsu {\it et al.} \cite{Miyatsu} who obtained neutron stars with masses in the range $1.8-2.1M_\odot$ using a chiral QMC model in the relativistic Hartree--Fock approximation when the SU(6) spin-flavor symmetry is relaxed to the SU(3)-flavor one. Dhiman {\it et al.} \cite{Dhiman07} found neutron star masses as large as $2.1M_\odot$ by varying the $\omega$-meson self-coupling and the hyperon-meson couplings in RMF models in such a way that the bulk nuclear observables, nuclear matter incompressibility coefficient, and hyperon-nucleon potential depths remain practically unchanged. Dexheimer and Schramm \cite{Dexheimer08} were also able to obtain neutron stars including hyperons with masses of $\sim 2.1M_\odot$ within a hadronic chiral SU(3) model.

Although the presence of hyperons in neutron stars seems to be energetically unavoidable, their strong softening effect on the EoS leads (except for the exceptional cases just mentioned) to maximum masses not compatible with observation. The solution of this problem requires a mechanism (or mechanisms) that could eventually provide the additional repulsion needed to make the EoS stiffer and, therefore the maximum mass compatible with the current observational limits. Three different mechanisms that could provide such additional repulsion have been proposed. They are: (i) the inclusion of a repulsive YY interaction through the exchange of vector mesons, higher order couplings or density dependent couplings 
\cite{glen,glenb,glenc,glend,rmf,rmfb,rmfc,rmfd,rmfe,rmff,rmfg,rmfh,Hofmann,Huber,Taurines,Gomes2014,Rikovska,Whittenbury,Thomas,Miyatsu,Dhiman07,Dexheimer08,Bednarek2011,Weissenborn1,Weissenborn2,Agrawal,Lopes2014,Oertel2014,Maslov,GuptaArumugam,CharBanik,Typel,Typelb,Colucci,vanDalen,Lim2014}, 
(ii) the inclusion of repulsive hyperonic three-body forces 
\cite{taka,takab,vidanatbf,yamamoto,yamamotob,yamamotoc,lonardoniprl}, or (iii) the possibility of a phase transition to deconfined quark matter at densities below the hyperon threshold
\cite{Alford07,WeissenbornSagert,Ozel,Schulze2011,SchrammActa,ZdunikHaensel,Klahn2013,Bonanno,Lastowiecki2012,Lastowiecki2015,Fraga2014,Masuda2013,Schramm1310}.
In the following we briefly review the first two solutions whereas the last one will be revised in more detail in section \ref{sec:qm_ns} after a couple of short comments on the role of the $\Delta$ isobar and  kaon condensation in neutron stars.

\begin{figure*}
\begin{center}
\resizebox{0.7\textwidth}{!}
{%
\includegraphics[clip=true]{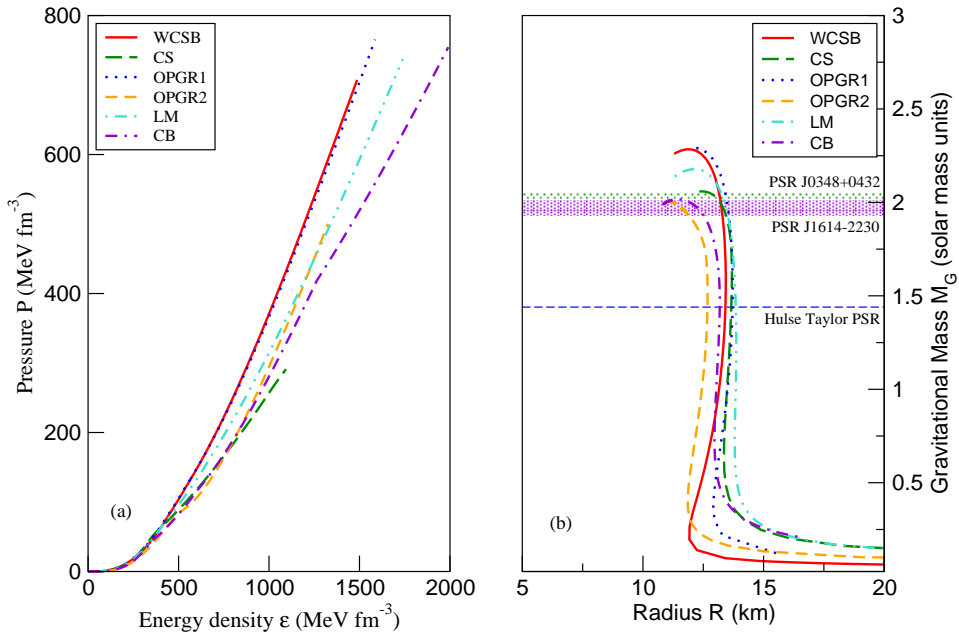}
}
\caption{(Color online)  Selected RMF EoSs (panel a) including hyperon-hyperon repulsion and their corresponding mass-radius relation (panel b) satisfying the $2M_\odot$ constraint. 
The horizontal lines and bands in panel (b) show the observational data of the Hulse--Taylor \cite{hulsetaylor,hulsetaylor2}, PSR J1614-2230 \cite{Demorest} and PSR J0348+0432 \cite{Antoniadis} pulsars.} 
\label{f:fig3}       
\end{center}
\end{figure*}

\begin{table}[b]
\begin{center}
\caption{Maximum masses and radii at 1.4 $M_{\odot}$ predicted by the 
selected models shown in Fig.\ \ref{f:fig3}.}
\label{tab:eosmr}       
\begin{tabular}{ l | c | c }
\hline\noalign{\smallskip}
\hline\noalign{\smallskip}
{EoS} & {$M^{max}$ ($M_{\odot}$)} & {$R_{1.4}$ (km)}  \\
\noalign{\smallskip}\hline\noalign{\smallskip}
{WCSB} & {2.28} & {13.4} \\
{CS}  & {2.06} & {13.7} \\
{OPGR1} & {2.29} & {13.8}\\
{OPGR2} & {2.01} & {12.7}\\
{LM} & {2.18} & {13.9} \\
{CB} & {2.02} & {13.2}\\
\noalign{\smallskip}\hline
\noalign{\smallskip}\hline
\end{tabular}
\end{center}
\end{table}
%

\begin{figure}[b]
\begin{center}
\resizebox{0.30\textwidth}{!}
{%
\includegraphics[clip=true]{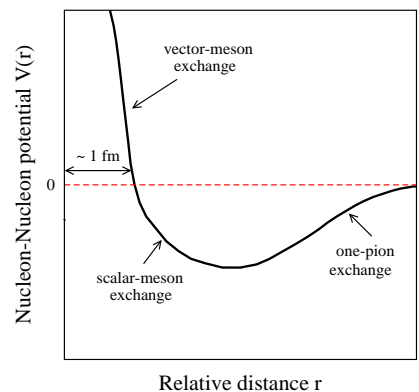}
}
\caption{(Color online) Schematic plot of the radial dependence of a generic nucleon-nucleon potential.} 
\label{f:fig2}       
\end{center}
\end{figure}

\subsection{Hyperon-hyperon repulsion}
\label{subsec:yvr}

This solution has been mainly explored in the context of RMF models. The number of works that have explored this solution to the hyperon problem in the last years is too large and, unfortunately, we cannot summarize all of them, and are forced to choose a few as representative of the copious research carried out. Consequently, we would like to apologize to those groups whose results are not included in this summary. Some of the selected EoSs including hyperon-hyperon repulsion and their corresponding mass-radius relation are shown, respectively, in panels (a) and (b) of Fig.\ \ref{f:fig3}. 
Maximum masses and radii at 1.4 $M_{\odot}$ predicted by these models are given in Table \ref{tab:eosmr}. The interested reader is referred to the original works for specific details.

As mentioned before, a repulsive YY interaction can be generated through the exchange of vector mesons, the inclusion of higher order couplings or the use of density-dependent couplings. The exchange of vector mesons is based on the well-known fact that,
in a meson-exchange model of nuclear forces,  vector mesons generate repulsion at short distances (see Fig.\ \ref{f:fig2}). If the interaction of hyperons with vector mesons is repulsive enough then it could provide the required stiffness to explain the current pulsar mass observations. However, hypernuclear data indicates that, at least, the $\Lambda$N interaction is attractive \cite{hashimoto06}. Therefore, in order to be consistent with experimental data of hypernuclei, the repulsion in the hyperonic sector is included only in the YY interaction through the exchange of the hidden strangeness $\phi$ vector meson coupled only to the hyperons. In this way, the onset of hyperons is shifted to higher densities and neutron stars with maximum masses larger than $2M_\odot$, and a significant hyperon content, can be successfully obtained. 

Several works have analyzed in detail this possibility. Bednarek {\it et al.} \cite{Bednarek2011}, for instance, proposed a non-linear RMF model involving hidden-strangeness scalar ($\sigma^*$) and vector ($\phi$) mesons, coupled only to hyperons and quartic terms involving vector meson fields in the the effective Lagrangian. These authors showed that the required stiffening necessary to allow neutron stars with hyperon cores and $M_{max} \geq 2M_\odot$ was in fact generated by the presence of the quartic terms involving the $\phi$ meson field. 

In a couple of recent works \cite{Weissenborn1,Weissenborn2}, one of the authors (D.C.) of this paper, performed a systematic study of the influence of the hyperon potentials within the RMF framework, and showed that the mass constraint could be reached through the inclusion of the $\phi$ meson mediating a repulsive interaction among hyperons, regardless of the uncertainties in their potentials. In particular, a detailed analysis of the influence of the $\phi$-meson-hyperon  coupling, going from the SU(6) quark model to a broader SU(3) symmetry was carried out in Ref.\ \cite{Weissenborn2} for various EoSs corresponding to different models. It was shown that the hyperon threshold is pushed to higher densities when the repulsion between hyperons induced by vector-meson exchange is gradually increased. The EoSs obtained in this work were quite stiff and
even if some of them give values of $M_{max}$ compatible with observation, however, they were not compatible with the data derived from the flow analysis of nuclear matter in heavy-ion reactions \cite{Danielewicz}. It is interesting to mention here that an extended RMF with parameters optimized to be compatible with selected nuclear observables such as binding energies and charge radii was employed by Agrawal {\it et al.} \cite{Agrawal}. The authors of \cite{Weissenborn2} also deduced that in order to be simultaneously consistent with the mass constraint and heavy-ion data, an EoS with hyperons requires larger hyperon-vector meson couplings. The role of the strangeness scalar meson $\sigma^*$, which induces attraction in the YY interaction, was neglected in this work following current experimental data on double-$\Lambda$ hypernuclei which suggest that the $\Lambda\Lambda$ interaction is very weakly attractive \cite{nagara}. A linear correlation between the maximum neutron star mass and the strangeness fraction in the core was proposed in this work. However, the contribution of the non-vanishing $\phi$ field was found (see the second entry in Ref.\ \cite{Weissenborn2}) 
to be erroneously omitted when determining the scalar-meson hyperon coupling
constants from the hyperon potential depths in nuclear matter at saturation, which gives a finite contribution to the cases studied beyond SU(6) symmetry to the more general SU(3) case. The main conclusions of this work remain, however, qualitatively unchanged, although, the relation between the maximum mass and the
strangeness content of the star could not be fit anymore by a simple linear formula as suggested in the
original work. This absence of a linear relation was also pointed out by the authors of Ref.\ \cite{Lopes2014}. The EoS and
the corresponding mass-radius relation that fulfills the $2M_\odot$ constraint in this model is shown by the
curves labelled WCSB in Fig.\ \ref{f:fig3}. This result corresponds to the case $z=0$, where $z$ is the ratio between the octet and the singlet baryon-vector meson coupling constants. 

Using the well-known quantum hadrodynamics, Lopes and Menezes \cite{Lopes2014} studied the effects of
the hyperon-meson coupling constants on the onset of hyperons in dense nuclear matter. They propose to use the SU(3)-flavor symmetry to fix the complete set of hyperon-meson couplings. The YY interaction was described only in terms of the $\phi$ vector meson. The models obtained were tested against experimental and astrophysical constraints obtained from heavy-ion collisions and neutron star phenomenology obtaining a maximum neutron star mass of $\sim 2M_\odot$. The curves labelled LM in Fig. \ref{f:fig3} show the EoS and the mass-radius relation obtained in this work to be compatible with observation.

An extensive parameter study of RMF models with hyperons have been performed very recently in Ref.\ \cite{Oertel2014}. As in similar works, the YY interaction was described in terms of the exchange of $\sigma^*$ and $\phi$ mesons. These authors found that it is possible to obtain high mass neutron stars with (i) a substantial amount of hyperons, (ii) radii of 12-13 km for the canonical mass of $1.4-1.5M_\odot$, and (iii) a spinodal instability at the onset of hyperons, if a sufficiently strong $\Lambda\Lambda$ interaction is assumed. Though this is in contradiction with experimental data, such inferences are yet inconclusive.
One of the EoS and the corresponding mass-radius relation satisfying  the $2M_\odot$ constraint in this model is shown by the curves labelled OPGR1  and OPGR1  in Fig.\ \ref{f:fig3}. These curves correspond, respectively, to the models GM1 Y6 and DDh$\delta$ Y6. For further details see Ref.\ \cite{Oertel2014}.

In a very recent work, Maslov {\it et al.} \cite{Maslov} have proposed a RMF model with hadron masses and coupling constants depending on the scalar meson field $\sigma$. All hadrons masses undergo a universal scaling, whereas the coupling constants are scaled differently. The model includes also  the  vector meson $\phi$ and hyperons.  The hyperon-vector-meson ($\omega, \rho,\phi$) coupling constants obey SU(6) symmetry relations. The model is flexible enough to satisfy constraints from heavy-ion collisions and astrophysical data.  The authors show that the hyperon puzzle can be partially solved if a reduction of the $\phi$ vector meson mass is taken into account.

All these works were based on the idea of generating YY repulsion through the exchange of vector mesons.
However, as it was mentioned above, YY repulsion can be also generated by introducing higher order couplings in the RMF model and/or density-dependent couplings.  In the following we comment briefly on three of the recent works based on these other possibilities.

A first example of these works is that of Bednarek {\it et al.} \cite{Bednarek2011} mentioned before. As
we said, these authors found that the presence of the quartic terms involving the $\phi$ meson field was required in order to get values for $M_{max}$ of neutron stars with hyperons compatible with observation.  

The role of higher order couplings, in this case in conjunction with kaon condensation, was also discussed
by Gupta and Arumugam \cite{GuptaArumugam}. These authors found that the higher order couplings play a significant role at higher densities, and that they  bring down the mass of a neutron star, which is further reduced in the presence of kaons to yield results consistent with the present observational constraints.

Char and Banik \cite{CharBanik} studied the effect of antikaon condensates in the presence of hyperons
in the framework of a RMF model with density-dependent couplings. The density dependence of nucleon-meson couplings were determined following the DD2 model of Typel {\it et. al} \cite{Typelb}.
The density-dependent hyperon-meson couplings were derived from the density-dependent nucleon-meson ones by using hypernuclear data, scaling laws and SU(6) symmetry. A repulsive YY interaction was also included by means of the exchange of $\phi$
mesons. The couplings for nucleon-antikaon interactions were obtained in a similar form. They obtained always maximum masses compatible with the current observational limits. The curves labelled CB in Fig.\ \ref{f:fig3} show one of the results of this work for the DD2 EoS of matter with nucleons, $\Lambda, \Xi^-$
and $\Xi^0$ with antikaon optical potential at zero momentum $U_{\bar K}=-140$ MeV.

Colucci and Sedrakian \cite{Colucci} carried out a systematic analysis of the hypernuclear matter EoS within the framework of relativistic energy density functional theory with density-dependent couplings. Having explored the parameter space of different hyperon-scalar meson couplings including mixing and breaking of the SU(6) symmetry, these authors arrived at a specific class of EoSs with weak \\hyperon-scalar meson couplings to describe massive stars with finite strangeness. This relativistic energy density functional was then used in Ref.\ \cite{vanDalen} to derive stringent constraints on the YN interaction by using combined 
hypernuclear data and the observational mass limits. The curve labels CS in Fig.\ \ref{f:fig3} show the results of this work for the values of the $\sigma$-hyperon coupling $x_{\sigma\Lambda}=0.6164$ and $x_{\sigma\Sigma}=0.15$. 

Before finishing this section, we would like to mention also the work of Lim {\it et al.} \cite{Lim2014}. These
authors employed Skyrme-type models and a finite-range force model to study the effect of the $\Lambda\Lambda$ interaction in neutron stars. The parameters of the model were fixed to reproduce single-particle energy levels and binding energies of double-$\Lambda$ hypernuclei.  They found that within this model neutron star structure depends strongly on the $\Lambda\Lambda$ interaction, and some of their parametrizations gave $M_{max} \geq 2M_\odot$.

As a final remark of this section, we would like to note that, although all these models are able to reconcile
the presence of hyperons in the neutron star interior with the existence of very massive neutron stars, one must still be a bit cautious. What these models do is basically a fit of several free parameters on the basis of our still quite scarce knowledge of the YY interaction. It is clear that neutron star observables can be used to better constrain the YY interaction; however, we will not have a hundred percent control on it until experimental data on YY scattering and multi-strange hypernuclei is available to check the validity of the
(still free) parameters derived and/or to complement them.

\begin{figure}[b]
\begin{center}
\resizebox{0.30\textwidth}{!}
{%
\includegraphics[clip=true]{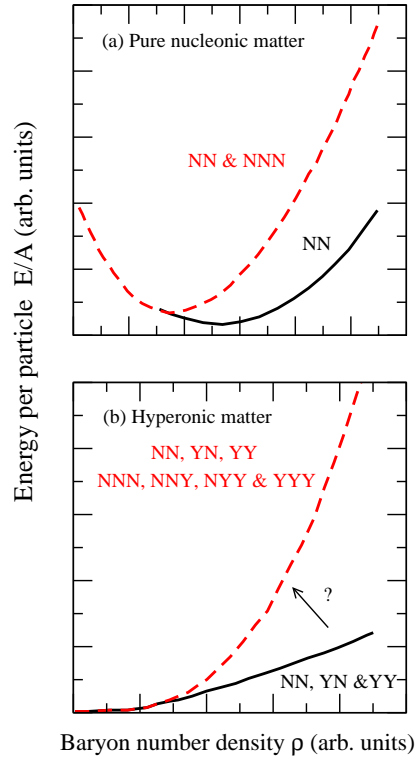}
}
\caption{(Color online) Qualitative illustration of the effect of three-nucleon and hyperonic three-body forces on the energy per particle of pure nucleonic (panel (a)) and hyperonic (panel (b)) matter.} 
\label{f:fig4}       
\end{center}
\end{figure}

\begin{figure*}[t]
\begin{center}
\resizebox{0.90\textwidth}{!}
{%
\includegraphics[clip=true]{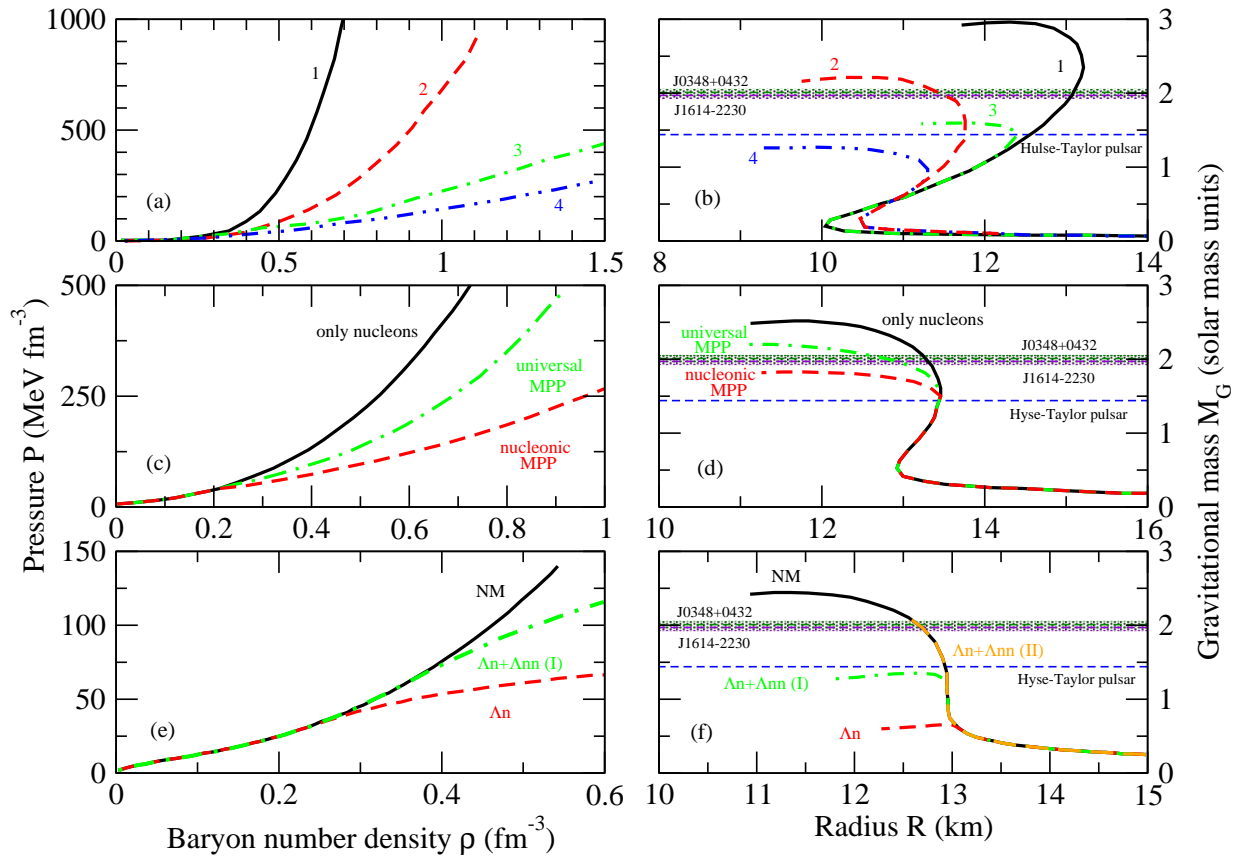}
}
\caption{(Color online). Summary of the equations of state and mass-radius relations obtained in Refs.\ \cite{vidanatbf} (panels (a) and (b)), \cite{yamamoto,yamamotob,yamamotoc} (panels (c) and (d)) and \cite{lonardoniprl} (panels (e) and (f)). The horizontal lines and bands in panels (b), (d) and (f) show the observational data of the Hulse--Taylor \cite{hulsetaylor,hulsetaylor2}, PSR J1614-2230 \cite{Demorest} and PSR J0348+0432 \cite{Antoniadis} pulsars.} 
\label{f:fig5a}       
\end{center}
\end{figure*}



\subsection{Hyperonic three-body forces}
\label{subsec:yyy}

It is well known that the inclusion of three-nucleon forces  in the nuclear Hamiltonian is fundamental to reproduce properly the properties of few-nucleon systems as well as the empirical saturation point of symmetric nuclear matter in calculations based on non-relativistic many-body approaches. Therefore, it seems natural to think that three-body forces (TBFs) involving one or more hyperons (NNY, NYY and YYY) could also play an important role in the determination of the neutron star matter EoS, and contribute to the solution of the hyperon puzzle. These forces could eventually provide, as in the case of the three-nucleon ones, the additional repulsion needed to make the EoS stiffer at high densities (see Fig.\ \ref{f:fig4}) and, therefore, make the maximum mass of the  star compatible with the recent observations. This idea was suggested even before the observation of neutron stars with $\sim 2M_\odot$ (see {\it e.g.,} Ref.\ \cite{taka,takab}), and it has been explored by some authors in the last years \cite{vidanatbf,yamamoto,yamamotob,yamamotoc,lonardoniprl}. In the next paragraphs we briefly revise the main results and conclusions of these works. A summary of these results is shown in Fig.\ \ref{f:fig5a}. The interested reader is referred to the original works in Refs.\ \cite{vidanatbf,yamamoto,yamamotob,yamamotoc,lonardoniprl} for the specific details of these calculations.

The authors of Ref.\ \cite{vidanatbf} have used a model based on a microscopic BHF approach of hyperonic matter
using the Argonne V18 \cite{av18} NN force and the Nijmegen YN soft-core NSC89 \cite{nijmegen} one supplemented with additional simple phenomenological density-dependent contact terms  to establish numerical 
lower and upper limits to the effect of hyperonic TBF on the maximum mass of neutron stars. 
Assuming that the strength of these forces is either smaller than or as large as the pure nucleonic ones, the results of this work show that although hyperonic TBF can reconcile the maximum mass predicted by microscopic approaches with the canonical value of $1.4-1.5M_\odot$, they are, however, unable to provide the repulsion needed to make the predicted maximum masses compatible with the recent observations of massive neutron stars. These results are summarized in panels (a) and (b) of Fig.\ \ref{f:fig5a} which show, respectively, the stiffer and softer
EoSs obtained for the pure nucleonic (curves 1 and 2) and hyperonic (curves 3 and 4) cases, and their corresponding mass-radius relation.
If the strength of hyperonic TBF is assumed to be larger than the the nucleonic one then, in principle, it is possible to obtain maximum masses of the order of $\sim 2M_\odot$ or larger. However, the authors of this work argued that this assumption gives rise to an EoS unrealistically stiff, the reason being the following: it is known that the strength of the two-body YN interaction is smaller than that of the NN one ({\it e.g.,} the single-particle potential of a $\Lambda$ in symmetric nuclear matter at saturation for zero momentum is about 1/3 that of the nucleons). Therefore, it seems quite natural to think that most probably the strength of hyperonic TBF is either smaller or as large as the pure nucleonic one, but not larger. This argument can seem a bit speculative. However, this work was exploratory, and its main aim was to serve as a motivation for more realistic and sophisticated studies of hyperonic TBF that are capable of giving a definite answer to this question.

Recently, Yamamoto {\it et al.} \cite{yamamoto,yamamotob,yamamotoc} have proposed a multi-Pomeron exchange potential (MPP) model to introduce universal three-body repulsion among three baryons (NNN, NNY, NYY and YYY) in the hyperonic matter EoS.
This universal three-body repulsive potential is based on the extended soft core (ESC) baryon-baryon interaction of the Nijmegen group \cite{nijmegen4,nijmegen5}. In addition, three-nucleon attraction (TNA) is added phenomenologically in order to reproduce the nuclear saturation properties precisely. The strength of the MPP is determined by analyzing the nucleus-nucleus scattering with the use of a $G$-matrix folding potential derived from the ESC interaction complemented with the MPP and the TNA. Some of the EoSs and mass-radius relations derived in this work are shown, respectively, in panels (c) and (d) of Fig.\ \ref{f:fig5a}. The black solid curves corresponds to the pure nucleonic case whereas the red dashed and green dot-dashed ones show the results including hyperons in addition to nucleons. The difference between the last two is that whereas the red dashed curves, as the black solid ones, include the contribution of the MPP only in the nucleonic sector the green dot-dashed curves include it also in the hyperonic one. As seen in panel (d), when the MPP contribution is included only in the nucleon sector, the maximum mass is reduced from 
$\sim 2.5M_\odot$ to $\sim 1.8M_\odot$ due to the presence of hyperons. However, when the MPP contribution is taken into account universally for all baryons, the maximum mass is recovered to $\sim 2.2M_\odot$, a value even larger that that of the recent observations. This is in contradiction with the results and conclusions of Ref.\ \cite{vidanatbf} where the case of a universal three-body repulsion was also analyzed concluding that even in this case
hyperonic TBFs are not enough to reconcile the predictions of microscopic calculations with observation. This is clearly still an open question that requires further analysis.

The first quantum Monte Carlo calculation of the EoS and the mass-radius relation of an infinite system of neutrons 
and $\Lambda$ hyperons have been performed also very recently by Lonardoni {\it et al.} \cite{lonardoniprl}. The calculation uses the Argonne V8' \cite{av8p} NN force and the Urbana IX \cite{uix} three-nucleon one together with a phenomenological $\Lambda$N interaction fitted to the available scattering data, and two different parametrizations of a $\Lambda$NN force which includes contributions from $s-$ and $p-$wave $2\pi$ exchange plus a
phenomenological repulsive term \cite{usmani}. The results of this calculation, summarized in panels (e) and (f) of Fig.\ \ref{f:fig5a},  show first that the onset of the $\Lambda$ hyperon in neutron matter (see Fig.\ 1 of Ref.\ \cite{lonardoniprl} ) depends strongly on the $\Lambda$NN force used. They indicate also that two parametrizations of the $\Lambda$NN force give very different results for the maximum mass.  One of them (orange dashed curve in panel (f)) gives a maximum mass compatible with $\sim 2M_\odot$. However, in this case the $\Lambda$ hyperon appears at a density larger than the one corresponding to the maximum mass. Therefore, in practice, this is a case in which in fact no $\Lambda$s are present in the neutron star interior. The authors conclude that, with the model they considered, the presence of hyperons in the core of neutron stars cannot be satisfactory established and, consequently, there is no clear incompatibility with astrophysical observations when $\Lambda$s are included. Using the own words of these authors, stronger experimental and theoretical constraints on the two- and three-body interactions involving hyperons are therefore necessary to properly assess the role of hyperons in neutron stars.

Before we finish this section we should also mention the recent DBHF calculation of the neutron star matter EoS with hyperons by Katayama and Saito \cite{dbhf2,dbhf2b}. It is known that in the DBHF approach of nuclear matter three-nucleon forces are partially included by means of nucleon-antinucleon virtual excitations in the scalar $\sigma$-meson exchange process due to the dressed Dirac spinor in the nuclear medium. Therefore, it is expected that hyperonic TBF can also be  partially included through the DBHF approach. The first extension of the DBHF approach to the hyperonic sector was done by Sammarruca in Ref. \cite{dbhf1} where this author studied the equation of state of symmetric nuclear matter with a moderate concentration of $\Lambda$ hyperons using YN potentials of the J\"{u}lich group \cite{juelich,juelich2}. However, neither the inclusion of other hyperon species nor neutron star matter was considered in this work. In this sense the calculation of Katayama and
Saito can be considered as the first DBHF calculation of neutron star matter with hyperons. These authors use the Bonn NN potential and impose SU(6) symmetry to fix the YN couplings. The calculation predicts a maximum mass of $\sim 2M_\odot$ compatible with the recent observations.

As we have just seen, at present it is still an open issue whether hyperonic TBFs can, by themselves, solve completely the hyperon puzzle or not. However, it seems that even if they are not the full solution, 
most probably they can contribute to it in an important way.

\subsection{$\Delta$ isobar and kaon condensation in neutron stars}
\label{subsec:delta}

An alternative way to circumvent the hyperon puzzle is to invoke the appearance of other hadronic degrees of freedom such as for instance the $\Delta$ isobar or meson condensates that push the onset of hyperons to higher densities. 

Usually, the $\Delta$ isobar is neglected in neutron stars since its threshold density was found to be higher than the typical densities prevalent in the neutron star core. However, this possibility has been recently reviewed by Drago {\it et al.} in Ref.\ \cite{Drago14}. The authors of this work have shown that the onset of the $\Delta$ depends crucially on the density-dependence of the derivative parameter of the nuclear symmetry energy, $L=3\rho_0(\partial E_{sym}(\rho)/\partial \rho)_{\rho_0}$. By using a state-of-the-art EoS and recent experimental constraints of $L$, these authors showed that the $\Delta$ isobar could actually appear before the hyperons in the neutron star interior. However, they found that, as soon as the $\Delta$ is present the EoS, as in the case of hyperons, becomes considerably softer and, consequently, the maximum mass is reduced to values below the current observational limit also in this case, giving rise to what has been 
recently called the $\Delta$ puzzle.

The possible existence of a Bose--Einstein condensate of negative kaons in the inner core of neutron stars has also been also been extensively considered in the literature (see {\it e.g.,} \cite{kaon1,kaon1b,kaon2,kaon3,kaon4,kaon5} and references therein). As the density of stellar matter increases, the $K^-$ chemical potential, $\mu_{K^-}$, is lowered by the attractive vector meson field originating from dense nucleonic mater. When $\mu_{K^-}$ becomes smaller than the electron chemical potential $\mu_{e}$ the process $e^-\rightarrow K^-+\nu_e$ becomes energetically possible. The critical density for this process was calculated to be in the range $2.5-5\rho_0$ \cite{kaon3,kaon4}. However, as in the case of the $\Delta$, the appeareance of the kaon condensation induces also a strong softening of the EoS and the consequently leads to a reduction of the maximum mass to values also below the current observational limits. The interested reader to the original works on this subject \cite{kaon1,kaon1b,kaon2,kaon3,kaon4,kaon5} for 
a comprehensive description of the implications of kaon condensation on the structure and evolution of neutron stars.

\begin{figure}[t]
\begin{center}
\resizebox{0.70\textwidth}{!}
{%
\includegraphics[clip=true]{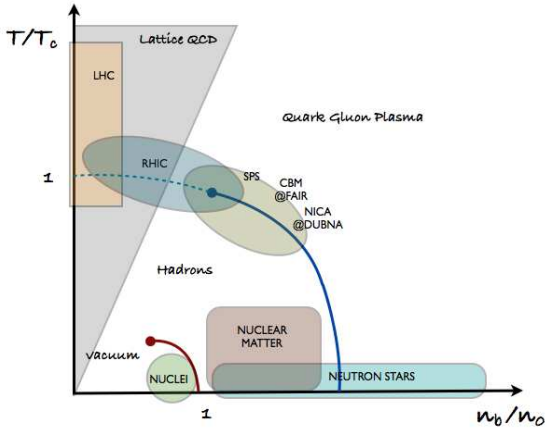}
}
\caption{(Color online) Schematic plot of the QCD phase diagram.} 
\label{f:fig5}       
\end{center}
\end{figure}

\section{Quarks in neutron stars}
\label{sec:qm_ns}

Strangeness is expected to appear in the interior of neutron stars also in a deconfined form. Compact stars which possess a quark matter core, either as a mixed phase of deconfined quarks and hadrons, or as a pure quark matter phase, surrounded by a shell of hadronic matter
are called hybrid stars. The description of these objects requires two EoSs describing, respectively 
the hadronic and quark phases. The phase transition between the two phases is usually described by means of the Maxwell or Gibbs constructions. The possible existence of a new class of compact stars completely made of deconfined $u, d,s$ quark matter (strange quark matter (SQM)), is one of the consequences of the Bodmer--Witten--Terezawa hypothesis \cite{sqm,sqm2,sqm3,sqm4,sqm5}, according to which SQM could be the true ground state of strongly interacting matter. These stars are usually called
strange stars.  

Current theoretical descriptions of quark matter at high density rely on phenomenological models, which are constrained using the few available experimental information on high density baryonic matter from heavy-ion collisions. A schematic plot of the QCD phase diagram is shown in Fig.\ \ref{f:fig5}. The figure illustrates the temperature and density regions explored by different theoretical approaches and several current or upcoming experimental facilities.  

It has been conjectured that quark matter in the core of neutron stars may be in the color superconducting state. There exist many models of color superconducting quark matter states (see {\it e.g.,} Ref.\ \cite{Alford08}). Of the many possible states, two important ones are the two-flavor color superconducting (2SC) state and the color flavor locked (CFL) superconducting one. In the 2SC state, only the light quarks $u$ and $d$ are paired, while in the CFL one all the three quarks are paired.

In the MIT bag model \cite{mit} (without invoking color superconductivity), a simple description of the quark phase deconfined from the hadronic phase is provided, in terms of the energy density of the quark gas, supplemented by a ``bag constant" that defines the energy difference between the perturbative vacuum and true vacuum. The Nambu-Jona-Lasinio (NJL) model \cite{njl} is a nonrenormalizable effective model valid upto a certain scale. It reproduces the origin of the nucleon mass analogous to the BCS theory of superconductivity and the origin of the constituent quark mass is related to the restoration and breaking of the chiral symmetry. However, due to its simplicity, it lacks some features of QCD such as the local color gauge symmetry. 

Different implications of deconfined quark matter in neutron stars will be addressed in several contributions
of this special issue (see {\it e.g.} \cite{dragoepja,dragoepja2,ignazioepja}), and we refer the interested reader to these works. In this section we are mainly interested and focus in the fact that an early deconfinement phase transition from hadronic mater to deconfined quark matter at densities below the hyperon threshold could provide a solution to the hyperon puzzle. The question that arises in this case is then whether quarks can provide the sufficient repulsion required to produce a $2M_\odot$ neutron star. Conversely, the observation of the 2$M_{\odot}$ neutron star may also help to impose important constraints on the models of hybrid and strange stars with a quark matter core, and improve our present understanding of the hadron-quark phase transition. Given below is a brief summary of some important conclusions drawn recently using the massive neutron star constraint. 

Back in 2007 following the discovery of the massive neutron star EXO 0748-676, Alford {\it et al.} \cite{Alford07} argued that hybrid or quark stars can reach a mass of 2$M_{\odot}$, as formerly demonstrated within the framework of MIT bag model, perturbative QCD as well as NJL models. 

Weissenborn {\it et al.} \cite{WeissenbornSagert} performed a systematic parameter study of the consequences of the neutron star mass limit on the properties of quark stars and hybrid stars with a pure quark core. Using an extended quark bag model, they concluded that massive strange stars require strong QCD corrections and large contribution from color superconductivity. For the case of hybrid stars, the EoSs with quark-hadron phase transition are compatible with the 2$M_{\odot}$ mass constraint, provided the quarks are strongly interacting and in color superconducting phase \cite{Ozel,WeissenbornSagert}. 

Within the framework of BHF theory, no heavy neutron stars with hyperons consistent with the mass constraint can be achieved, without invoking a transition to a deconfined quark phase \cite{Schulze2011}. A maximum mass  as large as 1.5$M_{\odot}$ can be obtained in the MIT bag model by employing an empirical density dependent bag constant, which is still much smaller than the observational constraint.

Schramm {\it et al.} \cite{SchrammActa} defined a hadronic flavor SU(3) model to combine hadronic and quark degrees of freedom in a unified way. They obtained a maximum mass of 2.06 $M_{\odot}$ for models with hyperons only, and reported a reduction of 10\% in the mass on inclusion of quark degrees of freedom. A further increase of about 20\% in the maximum mass was achieved by considering rotation. 

Zdunik and Haensel \cite{ZdunikHaensel} used the heavy neutron star mass observation to put general constraints on the 2SC and CFL phases of quark matter in the core of hybrid stars. They deduced that for thermodynamic stability, both a stiff hyperon repulsion at high baryon density accompanied by a stiff quark matter EoS are required to reproduce this result. 

Kl\"ahn {\it et al.} \cite{Klahn2013} described a large number of possible parametrizations based on the NJL model that can reach masses upto 2$M_{\odot}$. They demonstrated that to reconcile neutron star observations and heavy-ion flow data, large values of the diquark and scalar couplings are necessary as well as repulsion arising from the vector meson interaction. 

Bonanno and Sedrakian \cite{Bonanno} succeeded in constructing stable neutron star configurations with mass $\ge 2 M_{\odot}$ using an EoS exploiting features of phenomenological RMF density functional at low densities and the NJL model of superconducting quark matter at high densities. The constraints were satisfied subject to the condition that 
the nuclear EoS at post-saturation density is sufficiently stiff, followed by a transition to quark matter at a few times saturation
density, with substantial repulsive vector interactions in quark matter. 

Lastowiecki {\it et al.} \cite{Lastowiecki2012,Lastowiecki2015} studied the possibility of appearance of hyperons and strange quark matter in neutron stars subject to the constraints of observations of heavy compact stars
and flow constraints from heavy ion collisions \cite{Danielewicz}. Applying a color superconducting three-flavor NJL model for the quark sector and the DD2 model of \cite{Hofmann} and Dirac-Brueckner-Hartree-Fock model for the hadronic sector,
they showed that it is possible to have deconfined quark matter in the core of massive stars.

Using state-of-the-art techniques of perturbative QCD, Fraga {\it et al.} \cite{Fraga2014} constructed a simple
perturbative EoS of unpaired quark matter, as a possible alternative to the MIT bag model for free quark matter, and
applied it to successfully obtain pure quark stars with masses in excess of 2$M_{\odot}$. 

The nature of the hadron-quark phase transition has also been questioned in the works of Masuda {\it et al.} 
\cite{Masuda2013} and Schramm {\it et al.} \cite{Schramm1310}. Instead of assuming a first-order phase transition between hadronic and quark phases, governed
by Gibbs equilibrium conditions, they explored the possibility of a crossover at about $\sim 3\rho_0$.
They showed that this provides a novel mechanism to support massive neutron stars with a quark core, given that
the quark matter is strongly interacting in the crossover region and has a stiff EoS.

\begin{figure*}[t]
\begin{center}
\resizebox{0.70\textwidth}{!}
{%
\includegraphics[clip=true]{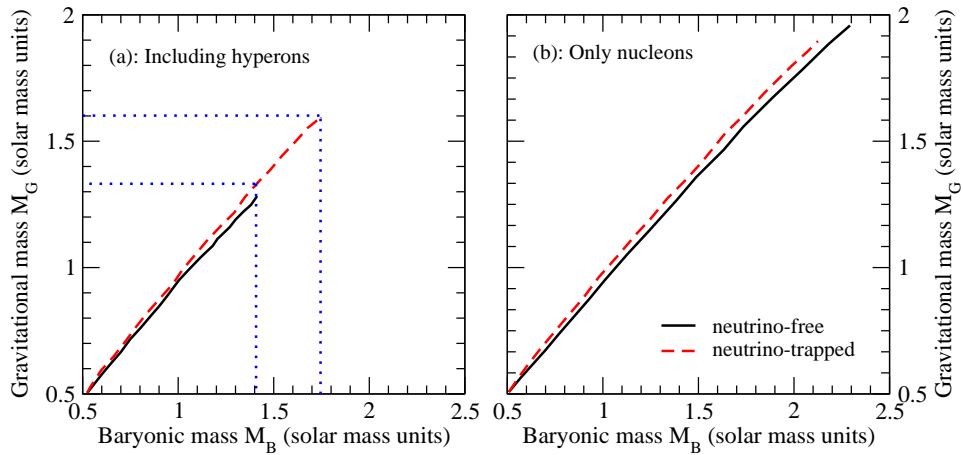}
}
\caption{(Color online) Gravitational mass as a function of the baryonic mass for neutrino-free (solid lines) and neutrino-trapped (dashed lines) matter. Panel (a) shows the results for matter containing nucleons and hyperons, whereas the results for pure nucleonic mater are shown in panel (b). Dotted horizontal and vertical lines show the window of metastability in the gravitational and baryonic masses. Figure adapted from Ref.\ \cite{vidanaAA}.} 
\label{f:fig6}       
\end{center}
\end{figure*}

\section{Hyperon stars at birth and neutron star cooling}
\label{sec:cool}

Neutron stars are formed after a type-II, Ib or Ic supernova explosion. Properties of
newly born neutron stars are affected with respect to NS ones by thermal effects and neutrino trapping. These two effects have a strong influence on the overall stiffness of the EoS and the composition of the star. In particular, see {\it e.g.,} \cite{keil,keilb,keilc,vidanaAA,burgio}, matter becomes more proton rich, the number of muons is significantly reduced, and the onset of hyperons is shifted to higher densities. In addition, the number of strange particles 
is on average smaller, and the EoS is stiffer in comparison with the cold and neutrino-free case.

A very important implication of neutrino trapping in dense matter is the possibility of having metastable  neutron stars
and a delayed formation of a ``low-mass'' ($M=1-2M_\odot$) black hole. This is illustrated in Fig.\ \ref{f:fig6} for the case 
of a BHF calculation of Ref.\ \cite{vidanaAA}. The figure shows the gravitational mass $M_G$ of the star as a function of its baryonic mass $M_B$. If 
hyperons are present (panel (a)), then deleptonization lowers the range of gravitational masses that can be supported by the EoS from about $1.59 M_\odot$ 
to about $1.28 M_\odot$ (see dotted horizontal lines in the figure). Since most of the matter accretion on the forming neutron star happens in a very early stages
after birth ($t<1$ s), with a good approximation, the neutron star baryonic mass stays constant during the evolution from the initial proto-neutron star configuration
to the final neutrino-free one. Then, for this particular model, proto-neutron stars which at birth have a gravitational mass between $1.28-1.59 M_\odot$ (a baryonic 
mass between $1.40-1.72 M_\odot$) will be stabilized by neutrino trapping effects long enough to carry out nucleosynthesis accompanying a type-II supernova 
explosion. After neutrinos leave the star, the EoS is softened and it cannot support anymore the star against its own gravity. The newborn star collapses then to a 
black hole \cite{keil,keilb,keilc}. On the other hand, if only nucleons are considered to be the relevant baryonic degrees of freedom 
(panel (b)), no metastability occurs and a black hole is unlikely to be formed during the deleptonization since the gravitational mass increases during 
this stage which happens at (almost) constant baryonic mass. If a black hole were to form from a star with only nucleons, it is much more likely to form during 
the post-bounce accretion stage.


The cooling of the newly born hot neutron stars is driven first by the neutrino emission from the interior,
and then by the emission of photons at the surface. Neutrino emission processes can be divided into slow and fast processes depending on
whether one or two baryons participate. The simplest possible neutrino emission process is the so-called direct Urca process
($n \rightarrow p+l+\bar \nu_l$, $p+l \rightarrow n +\nu_l$).  This is a fast mechanism which however, due to momentum conservation, it 
is only possible when the proton fraction exceeds a critical value $x_{DURCA} \sim 11\%$ to $15 \%$ \cite{lattimer}. Other neutrino processes
which lead to medium or slow cooling scenarios, but that are operative at any density and proton fraction, are the so-called modified Urca processes
($N+ n \rightarrow N+ p+l+\bar \nu_l$, $N+p+l \rightarrow N+n +\nu_l$), the bremsstrahlung ($N+N \rightarrow N+N + \nu +\bar \nu$), or
the Cooper pair formation ($n+n\rightarrow [nn]+\nu+\bar \nu$, $p+p\rightarrow [pp]+\nu+\bar \nu$), this last operating only when the
temperature of the star drops below the critical temperature for neutron superfluidity or proton superconductivity. If hyperons are present in the neutron star interior new 
neutrino emission processes, like {\it e.g.,} $Y\rightarrow B+l+\bar\nu_l$, may occur providing additional fast cooling mechanisms.  Such additional rapid cooling mechanisms, however, can lead to surface temperatures much lower than 
that observed, unless they are suppressed by hyperon pairing gaps. Therefore, the study of hyperon superfluidity becomes of particular interest since it 
could play a key role in the thermal history of
neutron stars. Nevertheless, whereas the presence of superfluid neutrons in the inner crust of neutron stars, and superfluid neutrons together with
superconducting protons in their quantum fluid interior is well established and has been the subject of many studies, a quantitative estimation of the
hyperon pairing has not received so much attention, and just few calculations exists in the literature \cite{super,super2,super3,super4,super5,super6,super7}.

\section{Hyperons and the r-mode instability of neutron stars}
\label{sec:rmode}

It is well known that the upper limit on the rotational frequency of a neutron star is set by its
Kepler frequency $\Omega_{Kepler}$, above which matter is ejected from the star's equator
\cite{lindblom86,lindblom86b}. However, a neutron star may be unstable against some perturbations which
prevent it from reaching rotational frequencies as high as $\Omega_{Kepler}$, setting, therefore,
a more stringent limit on its rotation \cite{lindblom85}. Many different instabilities can operate in
a neutron star. Among them, the so called r-mode instability \cite{anderson,andersonb}, a toroidal mode of
oscillation whose restoring force is the Coriolis force, is particularly interesting. This oscillation mode 
leads to the emission of gravitational waves in hot and rapidly rotating neutron stars though the 
Chandrasekhar--Friedman--Schutz mechanism \cite{cfs,cfs2,cfs3,cfs4}. Gravitational
radiation makes an r-mode grow, whereas viscosity stabilizes it. Therefore, an r-mode is unstable
if the gravitational radiation driving time is shorter than the damping time due to viscous processes.
In this case, a rapidly rotating neutron star could transfer a significant fraction of its rotational energy
and angular momentum to the emitted gravitational waves. These waves, potentially detectable, could
provide invaluable information on the internal structure of the star and constraints on the EoS.

Bulk ($\xi$) and shear ($\eta$) viscosities are usually considered the main dissipation mechanism of r-modes and other pulsation
modes in neutron stars. Bulk viscosity is the dominant one at high temperatures ($T> 10^9$ K) and, therefore,
it is important for hot young neutron stars. It is produced when the pulsation modes induce variations in pressure
and density that drive the star away from $\beta$-equilibrium. As a result, energy is dissipated as the weak
interaction tries to reestablish the equilibrium. In the absence of hyperons or other exotic components, the bulk
viscosity of neutron star matter is mainly determined by the reactions of direct  and modified  Urca processes. 
However, as soon as hyperons appear new mechanisms such as weak non-leptonic hyperon reactions 
($N+N\leftrightarrow N+Y$, $N+Y \leftrightarrow Y+Y$), direct ($Y \rightarrow B+l+\bar \nu_l$,
$B+l \rightarrow Y +\nu_l$) and modified hyperonic Urca ($B'+Y \rightarrow B'+B+l+\bar \nu_l$,
$B'+B+l \rightarrow B'+ Y +\nu_l$), or strong interactions ($Y+Y \leftrightarrow N+Y$, $N+\Xi \leftrightarrow Y+Y$, $Y+Y \leftrightarrow Y+Y$)
contribute to the bulk viscosity and dominate it for densities above $2-3 \rho_0 $. Several works have been devoted to the study of the hyperon bulk
viscosity \cite{ybv,ybv2,ybv3,ybv4,ybv5,ybv6,ybv7,ybv8,ybv9,ybv10,ybv11,ybv12,ybv13,ybv14,ybv15}. The interested reader is referred to these works for detailed studies on this topic. 

\begin{figure}[t]
\begin{center}
\resizebox{0.30\textwidth}{!}
{%
\includegraphics[clip=true]{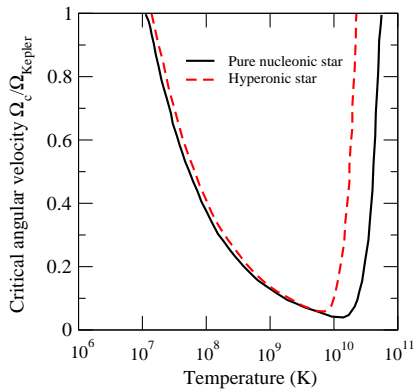}
}
\caption{(Color online) r-mode instability region for a pure nucleonic (solid line) and a hyperonic (dashed line) star with $1.27M_\odot$. The frequency of the mode has been taken as $\omega=10^4$ s$^{-1}$. Figure adapted from Ref.\ \cite{albertus}.} 
\label{f:fig7}       
\end{center}
\end{figure}

The time dependence of an r-mode oscillation is given by $e^{i\omega t-t/\tau}$, where $\omega$ is the frequency of the mode, and $\tau$ is an 
overall time scale of the mode which describes both its exponential growth due to gravitational wave emission as well as its decay due to viscous damping.
It can be written as $1/\tau(\Omega, T)=-1/\tau_{GW}+1/\tau_{\xi}+1/\tau_{\eta}$. If $\tau_{GW}$ is shorter than both $\tau_\xi$ and $\tau_\eta$ the mode will 
exponentially grow, whereas in the opposite case it will be quickly damped away. For each star at a given temperature T one can define a critical angular
velocity $\Omega_c$ as the smallest root of the equation $1/\tau(\Omega_c, T)=0$. This equation defines the boundary of the so-called r-mode instability region.
A star will be stable against the r-mode instability if its angular velocity is smaller than its corresponding $\Omega_c$. On the contrary, a star with 
$\Omega > \Omega_c$ will develop an instability that would cause a rapid loss of angular momentum through gravitational radiation if r-modes reach large amplitudes until its angular velocity
falls below the critical value. In  Fig.\ \ref{f:fig7} it is presented,  as example, the r-mode instability region for a pure nucleonic (solid line) and a hyperonic 
(dashed line) star with $1.27 M_\odot$ \cite{albertus}. 
The contributions to the bulk viscosity from 
direct and modified nucleonic Urca processes as well as from the weak non-leptonic process  $n+n\leftrightarrow p+\Sigma^-$ are included.
Clearly the r-mode instability window is smaller for the hyperonic star. The reason being simply  the increase of the bulk viscosity 
due to the presence of hyperons which makes the damping of the mode more efficient.

\section{Summary and future perspectives}
\label{sec:summary}

It is well known that the presence of hyperons in the neutron star interior, which is energetically favorable for densities above $2-3\rho_0$, leads to the softening of the neutron star EoS and to the 
reduction of $M_{max}$ to values that can be smaller than $2M_\odot$. Following the recent discovery of two millisecond pulsars with  masses of the order of $2M_{\odot}$ \cite{Demorest,Antoniadis}, the so-called ``hyperon puzzle", {\it i.e.,}  how to reconcile these measurements with the presence of hyperons in neutron stars, has become nowadays a 
subject of very active research and, particularly, the conditions under which hyperons appear in the neutron star core have been scrutinized in great detail by many authors. 

The presence of hyperons can be made compatible with the existence of massive neutron stars if there exist some mechanism (or mechanisms) that could provide at high densities the additional repulsion  
needed to make the EoS stiff enough for solving the hyperon puzzle. In this paper we have reviewed
some of the different mechanisms that have been proposed in the literature. The first of the mechanisms revised consists of the inclusion of a repulsive hyperon-hyperon interaction through the exchange of vector mesons, higher order couplings or density-dependent couplings
\cite{glen,glenb,glenc,glend,rmf,rmfb,rmfc,rmfd,rmfe,rmff,rmfg,rmfh,Hofmann,Huber,Taurines,Gomes2014,Rikovska,Whittenbury,Thomas,Miyatsu,Dhiman07,Dexheimer08,Bednarek2011,Weissenborn1,Weissenborn2,Agrawal,Lopes2014,Oertel2014,Maslov,GuptaArumugam,CharBanik,Typel,Typelb,Colucci,vanDalen,Lim2014}.
This possibility has been mainly explored in the context of RMF. However, as presently there is not enough experimental data to constrain the hyperon-nucleon and hyperon-hyperon interaction accurately, it is not clear whether the large repulsion invoked in such models is realistic. The second one requires the inclusion of repulsive hyperonic three-body forces
\cite{taka,takab,vidanatbf,yamamoto,yamamotob,yamamotoc,lonardoniprl}. However, at present it is still an open issue whether these forces can, by themselves, solve completely the hyperon puzzle or not, although, it seems that even if they cannot provide the full answer they can contribute to it in an important way. We also briefly discussed a possible solution to the hyperon puzzle by invoking the appearance of other hadronic degrees of freedom such as the $\Delta$ isobar or meson condensates that push the onset of hyperons to higher densities.  Finally, we showed that 
another way out of this dilemma is to consider the possibility of a phase transition to deconfined quark matter at densities below the hyperon threshold
\cite{Alford07,WeissenbornSagert,Ozel,Schulze2011,SchrammActa,ZdunikHaensel,Klahn2013,Bonanno,Lastowiecki2012,Lastowiecki2015,Fraga2014,Masuda2013,Schramm1310}.
However, the description of the quark phase via phenomenological models also suffer from uncertainties. Most models of hybrid stars unanimously agree that to construct massive neutron stars, they require both a sufficiently stiff hadronic EoS as well as a color superconducting quark phase with strong interaction among quarks to provide sufficient repulsion. 

If one makes a hasty conclusion, one may be prompted to think that as there exist several theoretical models which can 
successfully reproduce the 2$M_{\odot}$ observation, the ``hyperon puzzle" is rendered redundant. However, one must note that these models are able to achieve
the mass constraint by pushing the limits of their unknown parameters within the current range of uncertainties. Given the present
uncertainties in the models in absence of sufficient experimental or observational constraints, such models compatible with the mass observation 
are still permissible. It is however future facilities which will decide if such models are physically sound.
In short, in this article we have presented possible solutions to the hyperon puzzle within the current uncertainties, but in the absence of more restrictive experimental and observational
constraints the problem remains unsolved.

With the present lack of experimental  data  to constrain the uncertainties in theoretical models, one must look elsewhere for additional information. Several of the successful models have been tested against other constraints from nuclear and hypernuclear data as well as flow data in heavy-ion collisions. The price one must pay to invoke additional repulsion is a low strangeness content in the neutron star core, which should be compatible with other observables such as radii and cooling of neutron stars. Astronomical data could thus help to constrain the strangeness content and presence of exotica in the core. Oscillation modes in neutron stars are also another complementary tool to look for signatures of exotic matter in gravitational waves. 

A conclusive observation of multiply strange nuclear systems is absolutely necessary for a better understanding of the role of strangeness in neutron stars. The theories for the description of strangeness in massive neutron stars cannot be answered without the improved knowledge of $\Lambda\Lambda$ interaction, for which one requires careful high precision series of investigations of such an interaction. 

There are several new facilities planned or under construction such as in GSI in Germany, JLAB in USA and J-PARC in Japan. These facilities will hopefully provide much more precise updates on the properties of hyperon-nucleon and hyperon-hyperon interactions. Experimental hypernucleus physics is still an extremely active field of research.

In the next years, lattice QCD calculations will be able to provide, hopefully, the much required $\Lambda$N and $\Lambda$NN interactions \cite{Bedaque}. 
Recent lattice simulations of the binding energy of dibaryons containing hyperons indicate their existence.
Several of the hyperon EoSs compatible with the mass constraint go beyond the SU(6) symmetry towards a universal hyperon-meson coupling.
Whether such approaches are justified, will be clear only with experimental data of hyperon matter at large densities, such as from possible measurements of multi-hypernuclei at FAIR in GSI in the near future \cite{Gomes2014}. Further understanding of hyperons may be provided by the analysis of hyperon-hyperon correlation in heavy-ion collisions. 

Finally, new accurate measurements of masses and radii of neutron stars in future will play an important role in pinning down the EoS of nuclear matter at high densities. In addition, observational information such as cooling studies, pulsar back-bending, quasi-normal modes and gravitational waves would be useful tools to probe the existence of exotic matter in neutron stars.

\section*{Acknowledgements}

The authors are very grateful to S. Banik,  P. Char, L. L. Lopes, D. P. Menezes, M. Oertel, A. Sedrakian and S. Weissenborn for providing the data of their results shown in Fig.\ \ref{f:fig3}, and to D. Unkel for helpful and stimulating discussions. This work is partly supported by the project PEst-OE/FIS/UI0405/2014 developed under the inititative QREN financed by the UE/FEDER through the program COMPETE-“Programa Operacional Factores de Competitividade”, and by “NewCompstar”, COST Action MP1304.



\end{document}